\documentclass[useAMS,usenatbib]{mn2e}

\usepackage[dvips]{graphicx}
\usepackage{amssymb}
\newcommand{\aap}{A\&A}
\newcommand{\aaps}{A\&AS}

\newcommand{\aj}{AJ}
\newcommand{\apj}{ApJ}
\newcommand{\apjl}{ApJL}

\newcommand{\mnras}{MNRAS}

\newcommand{\pasp}{PASP}

\title[Photometric variability of WC9 stars]{Photometric variability of WC9 stars}
\author[R. Fahed, A.F.J. Moffat, A.Z. Bonanos]{R\'emi Fahed $^{1}$\thanks{fahed@astro.umontreal.ca }, Anthony F.J. Moffat $^{1}$\thanks{moffat@astro.umontreal.ca } and Alceste Z. Bonanos$^{2}$\thanks{bonanos@dtm.ciw.edu }\\
$^{1}$D\'epartement de physique, Universit\'e de Montr\'eal, C.P. 6128, Succ. C.-V., Montr\'eal, QC, H3C 3J7,\\ and Centre de recherche en astrophysique du Qu\'ebec, Canada\\
$^{2}$Carnegie Institution of Washington, Department of Terrestrial Magnetism, 5241 Broad Branch Road, NW, Washington, DC 20015}
\begin{document}

\date{}

\pagerange{\pageref{firstpage}--\pageref{lastpage}} \pubyear{2002}

\maketitle

\label{firstpage}

\begin{abstract}
Do some Wolf-Rayet stars owe their strong winds to something else
besides radiation pressure? The answer to this question is still not
entirely obvious, especially in certain Wolf-Rayet subclasses, mainly
WN8 and WC9. Both of these types of Wolf-Rayet stars are thought to be
highly variable, as suggested by observations, possibly due to
pulsations. However, only the WN8 stars have so far been vigorously and
systematically investigated for variability. We present here the results
of a systematic survey during 3 consecutive weeks of 19 Galactic WC9
stars and 1 WC8 star for photometric variability in two optical bands, V
and I. Of particular interest are the correlated variations in
brightness and colour index in the context of carbon-dust formation,
which occurs frequently in WC9 and some WC8 stars. In the most variable
case, WR76, we used this information to derive a typical dust grain size
of $\sim$ 0.1 $\mu$m. However, most photometric variations occur at
surprisingly low levels and in fact almost half of our sample shows no
significant variability at all above the instrumental level ($\sigma
\sim$ 0.005 -- 0.01 mag).
\end{abstract}


\section{Introduction}
Population I Wolf-Rayet (WR) stars are the descendants of the most
massive stars on the main sequence, namely O stars. They present the
signatures of strong, dense winds, in which typical terminal velocities
are of order 10$^3$ km s$^{-1}$. WR spectra come in two sequences, those
with strong lines of helium and nitrogen (WN subtypes) and those with
strong lines of helium, carbon and oxygen (WC and WO). WC9 are the
coolest stars among population I WC subtypes, with spectroscopic
temperatures of order 50000 K at the stellar (hydrostatic) radius,
deduced from non-LTE, clumped, line-blanketed models \citep{Dessart2000,
Crowther2002, Barniske2006}. The particularity of WC9 stars is their
frequent very bright luminosity in the infrared that is associated with
dust formation \citep{Williams1987}.  Some WC9 stars also show strong
spectroscopic and photometric variability \citep{Crowther1997,
Eenens2003, Kato2002, Lepine1996}. The brightest WC9 star, WR103, was
observed to show periodic behaviour, suggesting a possible binary
\citep{Isserstedt1981, Moffat1986}, although later more intense data
\citep{Veen1999} suggest that stellar pulsations are at play, as also
confirmed by recent MOST satellite data \citep{Moffat2008}.  A few other
WC9 stars have been monitored in the optical by \citet{Veen2000},
revealing behaviour that reminds one of the all-variable WN8 stars,
which typically shows a level of photometric variability of $\sigma
\simeq 0.016$ to $ 0.05 $ mag in the visible
\citep{Lamontagne1987,Antokhin1995,Marchenko1998,Lefevre2005}. Photometric
variability of other WC9 stars both in the infrared and in the visible
can be explained by episodic or periodic dust formation (NIR emission
and optical absorption) associated in most cases with binarity
\citep{Veen1998,Williams1996,Williams2000,Williams2005}.

The goal of this study is to provide a systematic, uniform survey of
WC9-stars using precision optical photometry in an attempt to answer the
questions raised by previous studies, namely: Do all WC9 stars show
variability like that of WN8 stars? and Is variability associated with
dust formation or with pulsations or both?

\section{Observations}

We selected a complete sample of 19 WC9 stars and 1 WC8 star in the
southern sky from the ``VIIth Catalogue of Galactic Wolf-Rayet Stars''
and its annex \citep{vanderHuchtCatalog,vanderHuchtCatalogA} with V-band
magnitude from 12 to 16, in order not to go below the minimum practical
exposure time of the CCD and to stay below a reasonable maximum exposure
time of 600 seconds. Observations were made using the Swope 1-meter
telescope at Las Campanas (Chile) between UT 2007 June 8 and July 7. We
used the direct CCD camera SITe\#3 and the two optical bands V and I in
order to be sensitive to both magnitude and colour variations, which are
of particular interest to constrain the local dust properties. Our field
of view is 8.5$\times$10 arcminutes with a pixel size of 0.435". We
chose net exposure times such that we obtained a good signal-to-noise
ratio (at least $\sim$300) per observation with one or two observations
per clear night for each star. Standard star fields were not observed,
since we are only interested in differential photometry.

\section{Data reduction}
Extraction of stellar magnitudes from the images was made in the usual
way, using nightly flat-field images in each band, nightly bias images
and overscan pixels in each image.  We then chose to perform aperture
photometry on our images, which is more efficient and precise than the
alternative point-spread extraction for well-isolated stars in the
field. We first interpolated all the images to match a reference frame
so that the target star always has the same pixel position. This was
done with the ISIS routine interp.csh \citep{Alard2000}. We then chose
an adequate number of reference stars ($N_{Ref}$) for which we measured
the instrumental flux $F$. To do so, we used the IDL function APER with
a fixed aperture radius $R=6$ pixels (2.610'') in I band and $R=8$
pixels (3.480'') in V band, which correspond to 2$\times$FWHM for
typical images, together with a sky annulus of radii $[1.5\times
R,2.5\times R]$ . The instrumental flux is then computed as
\begin{displaymath}
F=\sum_{r < R} P(r) - \pi R^2 \times F_{sky}
\end{displaymath}
where P(r) is the pixel value at radius r and $F_{sky}$ is the mean flux
 computed in the sky annulus.  The differential magnitude $\Delta M_i$
 of each star in the field is finally given by :
\begin{displaymath}
 \Delta M_i = -2.5\,log\left( \frac{F_i}{\sum_{j\neq i}^{N_{Ref}} F_j} \right) 
\end{displaymath}
We then sorted the stars in order of variance and chose the $N_{Ref}$
least variable among them to be the new reference stars. We then
re-performed the process until it converged. $N_{Ref}$ is chosen in
order to minimize the variance of the reference stars and ranges from 10
for a very diffuse field to 50 or more for a crowded field. The
photometric precision reached goes from $\sigma(\Delta M) =$ 0.005 to
0.012 magnitude depending on the quality of the images, essentially
determined by the average airmass at which the star could be observed.

\section{Results}

\begin{figure*}
  \centering
  \includegraphics[width=17cm]{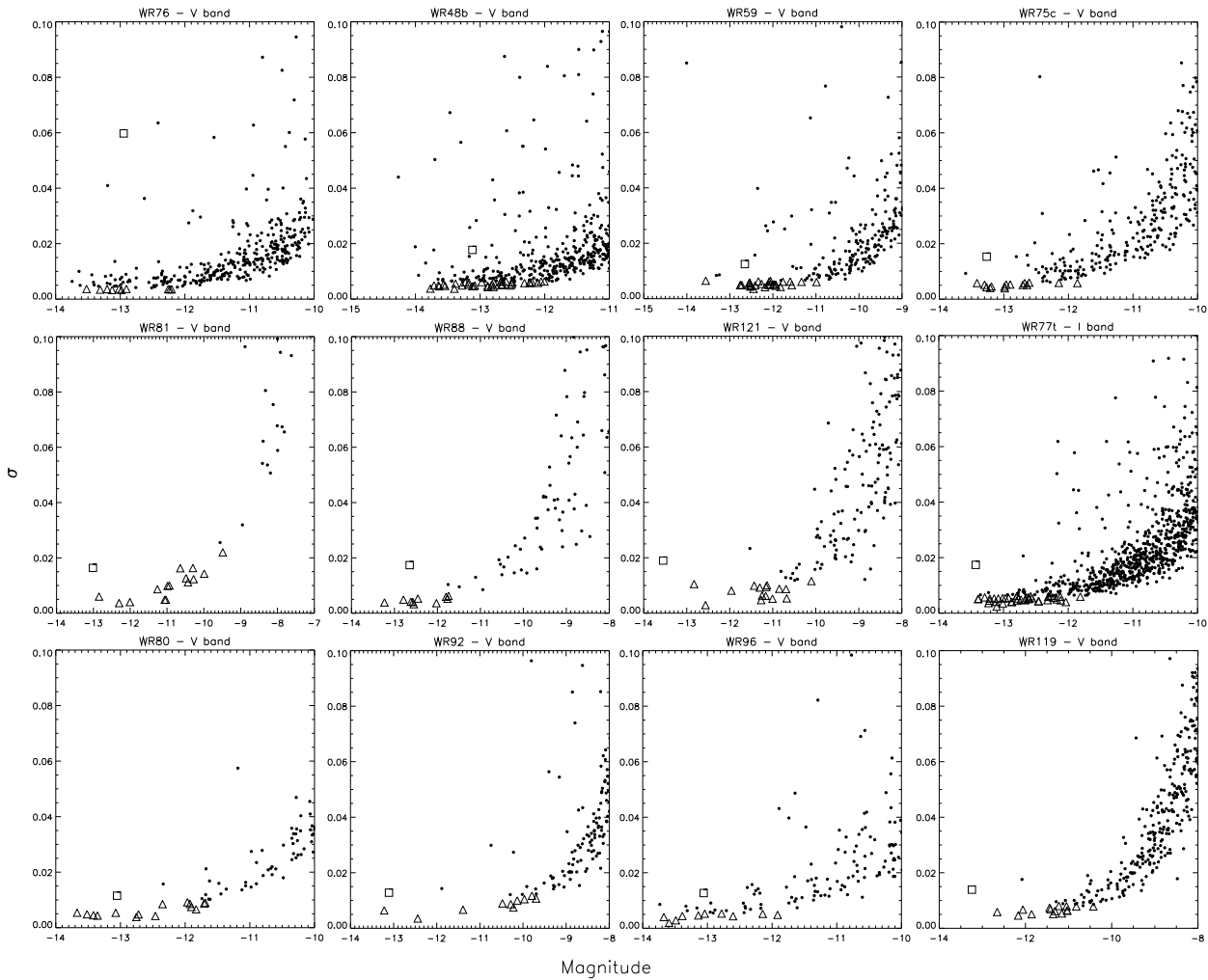}
  \caption{Dispersion plots for the fields containing the 12 variable
  WC9 stars: WR76, WR48b, WR59, WR75c, WR81, WR88, WR121, WR77t, WR80,
  WR92, WR96 and WR119. These are for the V band except for WR77t, which
  has been observed only in I band. The reference stars are indicated by
  triangles and the Wolf-Rayet by a square. On the x-axis is the
  instrumental magnitude of the stars measured with arbitrary zero point
  from an image chosen as reference. On the y-axis is the standard
  deviation in magnitudes of $\Delta M$ with time ($\sigma$).}
  \label{VariableStars}
\end{figure*}

\begin{figure*}
  \centering
  \includegraphics[width=17cm]{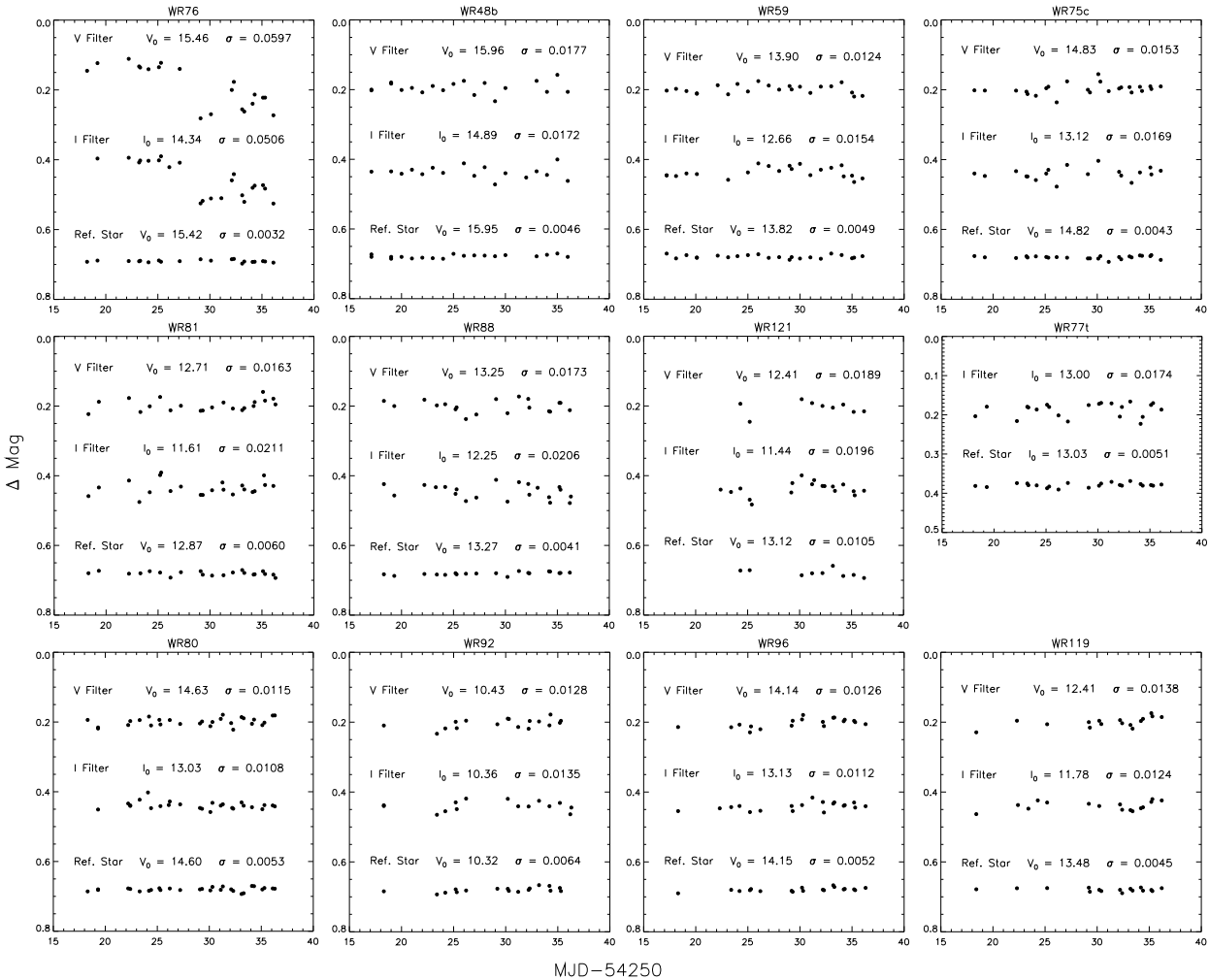}
  \caption{V and I lightcurves for the 12 variable WC9 stars compared
  with a reference star of approximately the same magnitude in
  V. $\sigma$ is the rms standard deviation of $\Delta Mag$ in
  time. When available, $V_0$ and $I_0$ are the apparent magnitudes of
  the stars from \citet{vanderHuchtCatalog,vanderHuchtCatalogA} . When
  not available, $V_0$ and $I_0$ are estimated values deduced from an
  extrapolation of what was available in the catalogue. MJD is for
  modified Julian day.}
  \label{VariableLC}
\end{figure*}

We found 12 WC9 stars to be variable when compared to reference stars of
similar magnitude : WR76, WR48b, WR59, WR75c, WR81, WR88, WR121, WR77t,
WR80, WR92, WR96 and WR119 (Fig.\ref{VariableStars} and
\ref{VariableLC}). Their variability ranges from $\sigma =$ 0.012 to
0.06 mag, the last value referring to the by far most variable star of
the sample, WR76.  Aside from this star, the level of variability of our
sample is quite small (going from $\sigma$ 0.012 to 0.02 mag) relative
to the "violent" WR103 described by \citet{Veen1999}, who confirm a
value of $\sigma(\Delta M)$ of $\simeq$0.035 mag in V band (0.05 mag
peak to valley) initially found by \citet{Isserstedt1981} and confirmed
by \citet{Moffat1986} and \citet{Balona1989}.  The timescale of the
(stochastic) variability of WR76 lies in the range of $\sim$ a day to
several days, like that seen in WR103.

\begin{figure*}
  \centering
  \includegraphics[width=17cm]{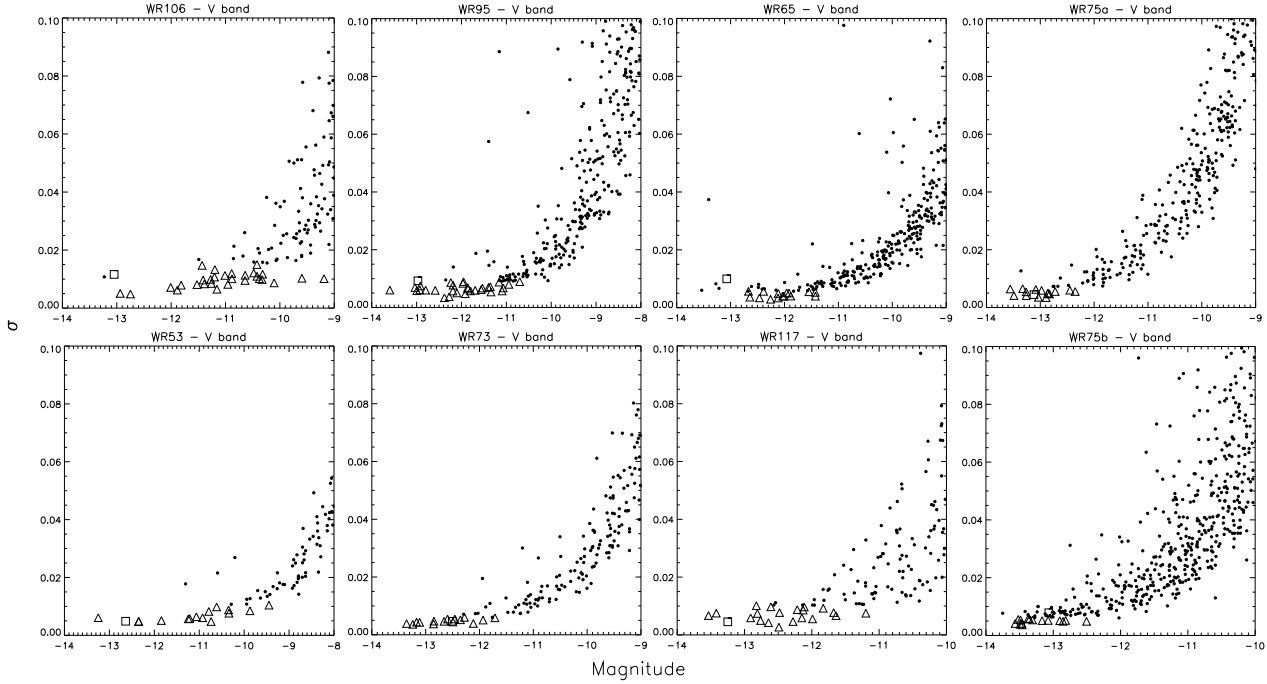}
  \caption{Dispersion plots for the fields containing the 7 non-variable
  WC9 stars WR106, WR95, WR65, WR73, WR75a, WR75b and WR117, and the
  non-variable WC8 star WR53 as in Fig.4. These are all for the V band.}
  \label{NonVariableStars}
\end{figure*}

\begin{figure*}
  \centering
  \includegraphics[width=17cm]{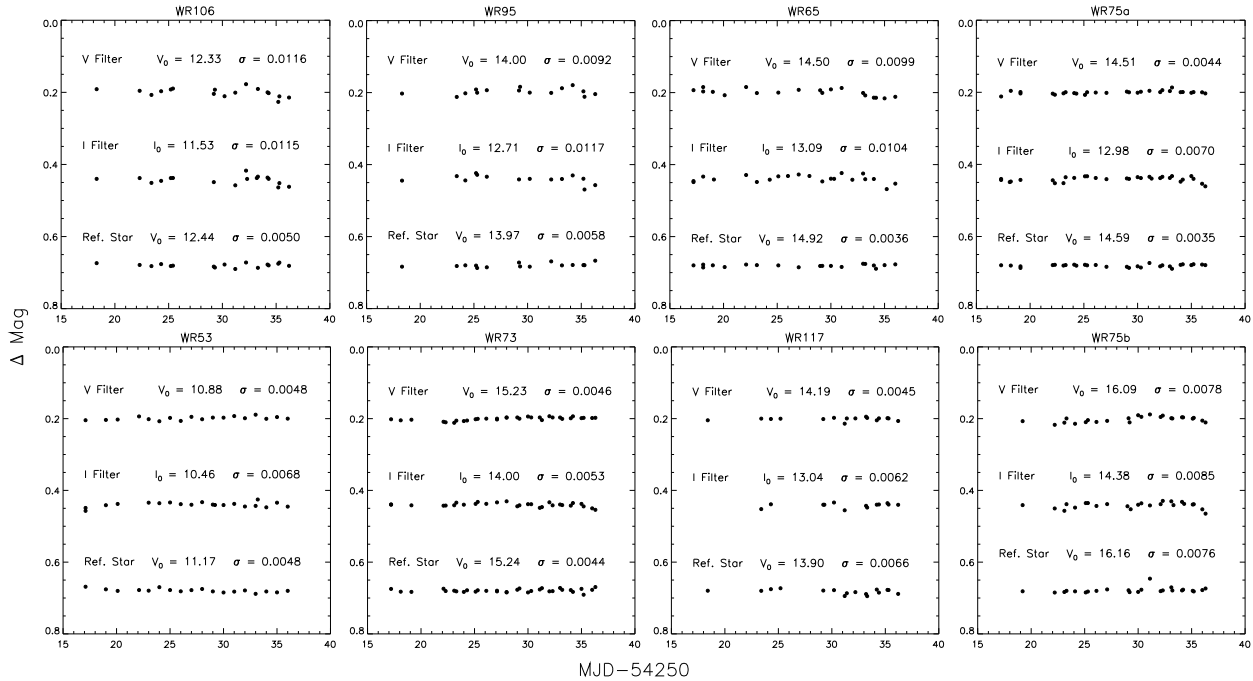}
  \caption{V and I lightcurves for the 8 non-variable WC9 stars compared
  with a reference star of approximately the same magnitude in V .}
  \label{NonVariableLC}
\end{figure*}

Eight stars were not detected as variables above the instrumental value
(Fig.\ref{NonVariableStars} and \ref{NonVariableLC}). Among these stars
we estimated the upper bound for $\sigma_V$ to be around 0.01 for WR106,
WR95 and WR65 and around 0.005 for WR53, WR73, WR75a, WR75b and WR117.

\begin{table*}
   \centering
  \begin{tabular}{|c|c|c|c|c|c|c|}
   \hline
 Star name & Spectral Type & $\sigma_V$ of WR & $\sigma_V$ of typical reference star & $\sigma_I$ of WR & $\sigma_I$ of typical reference star & Slope \\
\hline
WR76 & WC9d & 0.060 & 0.004 & 0.051 & 0.006 & 4.47$\pm$0.60 \\
WR48b & WC9d & 0.018 & 0.005 & 0.017 & 0.005 & - \\
WR59 & WC9d & 0.012 & 0.005 & 0.017 & 0.007 & - \\
WR75c & WC9 & 0.015 &  0.005 & 0.017 & 0.005 & - \\
WR81 & WC9 & 0.016 & 0.010 & 0.021 & 0.010 & - \\
WR88 & WC9 & 0.017 & 0.005 & 0.021 & 0.006 & - \\
WR121 & WC9d & 0.019 & 0.008 & 0.020 & 0.009 & - \\
WR77t & WC9d & - & - & 0.017 & 0.005 & - \\
WR80 & WC9d & 0.012 & 0.006 & 0.011 & 0.007 & - \\
WR92 & WC9 & 0.013 & 0.008 & 0.014 & 0.007 & - \\
WR96 & WC9d & 0.013 & 0.004 & 0.011 & 0.007 & - \\
WR119 & WC9d & 0.014 & 0.006 & 0.012 & 0.005 & - \\
WR106 & WC9d & 0.012 & 0.010 & 0.012 & 0.008 & - \\
WR95 & WC9d & 0.009 & 0.007 & 0.012 & 0.005 & - \\
WR65 & WC9 & 0.001 & 0.004 & 0.010 & 0.007 & - \\
WR75a & WC9 & 0.004 & 0.005 & 0.007 & 0.007 & - \\
WR53 & WC8d & 0.005 & 0.007 & 0.007 & 0.006 & - \\
WR73 & WC9d & 0.005 & 0.005 & 0.005 & 0.005 & - \\
WR117 & WC9d & 0.005 & 0.007 & 0.006 & 0.008 & - \\
WR75b & WC9 & 0.008 & 0.005 & 0.009 & 0.005 & - \\
\hline
  \end{tabular}
\caption{Measured V and I variability compared to typical values for the
reference stars. The latter is estimated from the rms mean of the
reference stars $\sigma$. 'Slope' is the slope of the (V) vs (V-I) plot
(Fig. 1), which was only significant for the most variable star of the
survey, WR76.}
\label{WRinfo}
\end{table*}

\begin{figure}
  \includegraphics[width=\columnwidth]{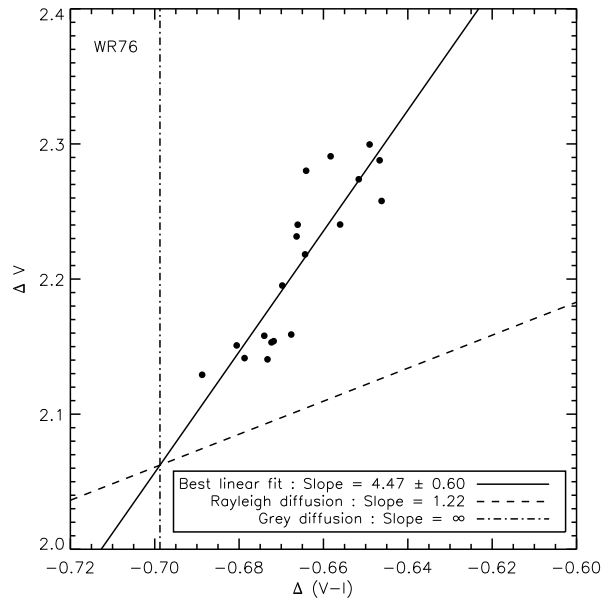}
  \caption{V band differential magnitude versus V-I differential colour
  index for WR76, by far the most variable star of our survey. The
  expected curves for Rayleigh scattering (valid for dust grains smaller
  than $\lambda_V/10$ $\simeq$ 0.05 $\mu$m) and grey scattering (valid
  for dust grains larger than $\lambda_V$ $\simeq$ 0.5 $\mu$m) are shown
  for comparison. The Rayleigh slope has been computed for Mie
  scattering with $Q_{ext}(\lambda,a) \propto \lambda^{-4}$ (see text).}
  \label{CorrelWR76}
\end{figure}

\section{Dust formation}

\begin{figure}
  \includegraphics[width=\columnwidth]{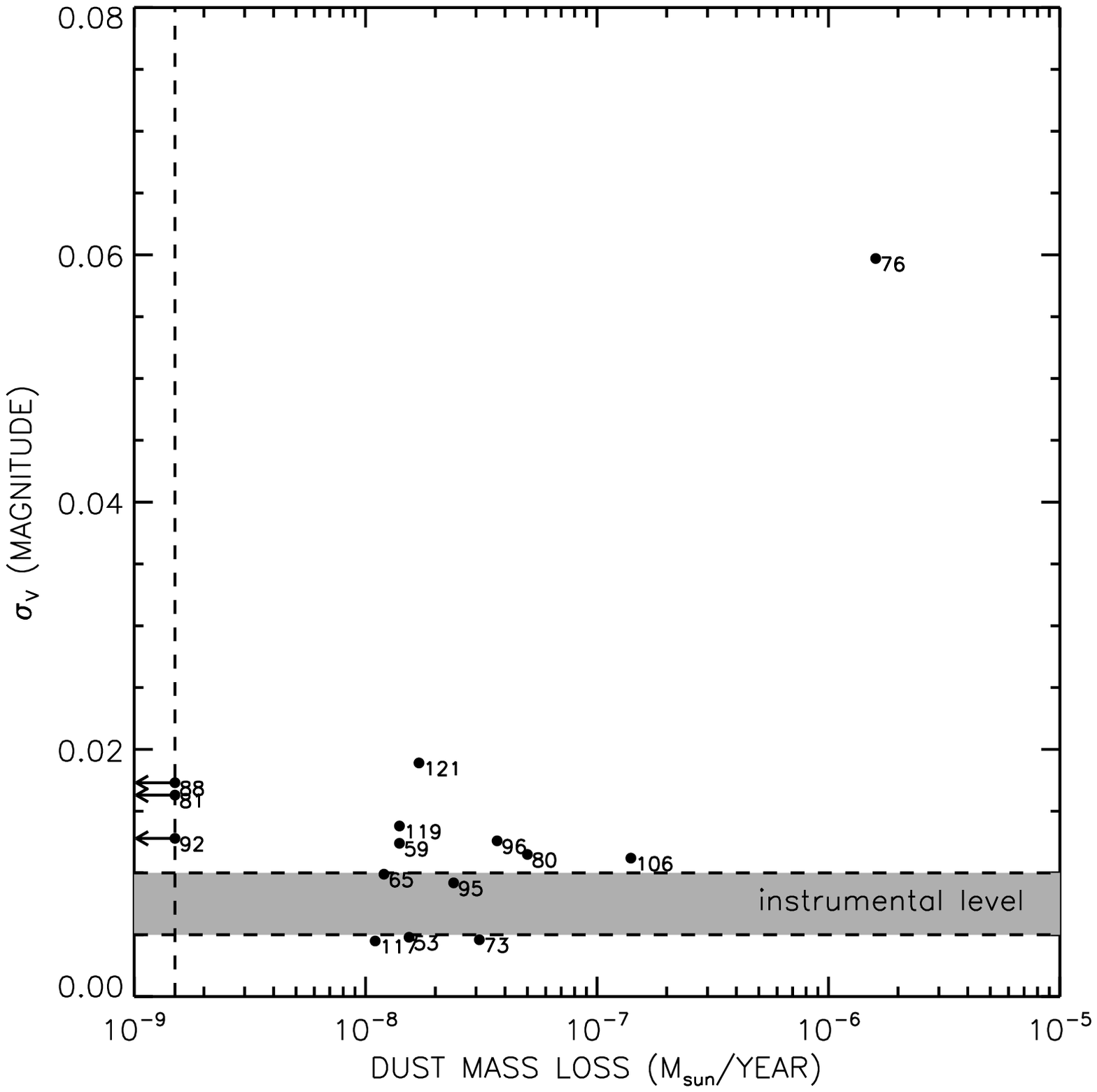}
  \caption{Our measured level of photometric variability $\sigma (V)$
  versus the estimated dust formation rate by \citet{Williams1987}. We
  show the range of values for the instrumental level, which is
  determined, for each field, by the $\sigma$ of the reference star of
  closest magnitude to the WC9 star. The dust formation rate estimates
  are based on infrared photometry. We see that any correlation between
  the level of variability and the dust formation rate of the WC9 stars
  is completely dominated by only one star, WR76, combined with a cloud
  of points at low variability level.}
  \label{sigmaVSmdotdust}
\end{figure}

We searched for a correlation of V vs. (V-I) but only found one for
WR76, which is by far the most variable star of our survey
(Fig.\ref{CorrelWR76}).  The computed correlation coefficient for this
star is R=0.92 and the Spearman rank index, which expresses the
probability that the two variables are uncorrelated, is $2.68\times
10^{-8}$ . We observe that WR76 becomes redder when the V extinction
increases (the star gets fainter).  This is consistent with
circumstellar extinction by small dust grains, which is more efficient
in V band than in I band. Moreover, WR76 is known to have a very high
dust formation rate measured from infrared photometry
\citep{Williams1987}: its dust formation rate is by far the highest
among our sample. Nevertheless, the correlation between the dust
formation rate and the level of variability of the star is completely
dominated by this star, as seen in Fig.\ref{sigmaVSmdotdust}. For the
remaining stars, the correlation is quite poor. In particular, we found
variable stars of low amplitude with no dust formation detected (WR75c,
WR81, WR88, WR92) together with non variable stars known to produce dust
because of their strong infrared excess (WR53, WR73, WR117). This might
be at least partly due to the shorter timescale of our variability
search compared to that for typical dust ejection.

As some non-dusty WC9 stars show low-level variability, presumably some
of that observed from WR76 may also be intrinsic. However, on account of
its high dust formation rate, variable extinction is thought to play a
dominant role.

Estimations of the dust formation rates were not available for all our
target stars, although the analysis of this correlation does show quite
clearly that the production of dust cannot explain the behaviour of all
WC9 stars. Like in the strongly variable WC9 star WR103
\citep{Veen1999,Moffat2008}, we believe that pulsations are likely to
play an important role in some cases, sometimes combined with the
effects of dust formation.

\section{Dust properties}

\begin{figure}
  \includegraphics[width=\columnwidth]{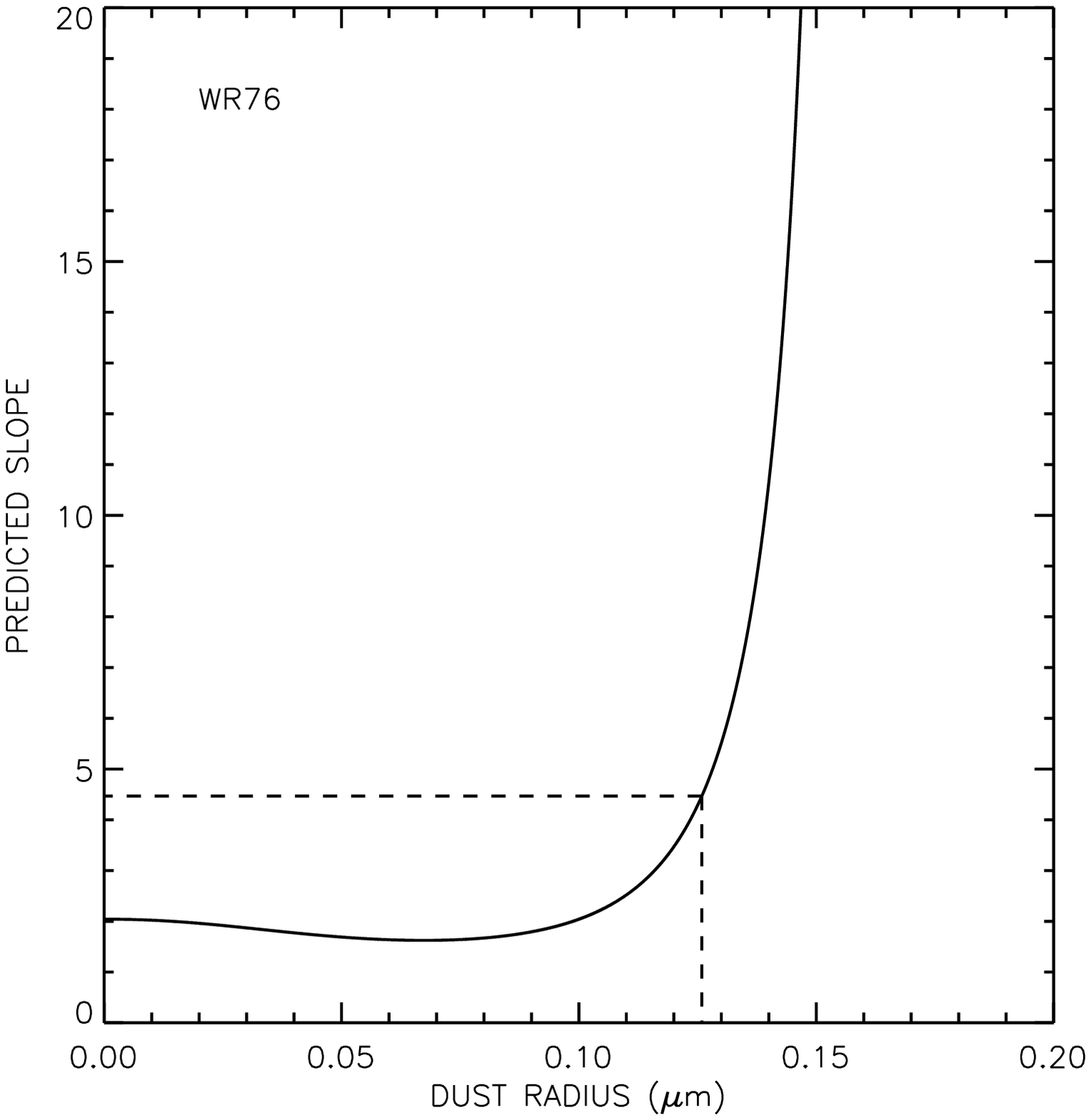}
  \caption{Model prediction of the V vs. (V-I) slope for WR76. We used
  Mie theory for mono-sized spherical grains plus optical constants for
  amorphous carbon by \citet{Zubko1996}. The best model to reproduce our
  slope of $4.47 \pm 0.60$ is a dust size of $a=0.126 \pm 0.003\, \mu
  m$}
  \label{dustsizeplot}
\end{figure}

It is known that dust in WC9 stars is essentially composed of amorphous
carbon grains \citep{Williams1987,Zubko1998,Marchenko2002}. We therefore
used Mie theory \citep{Bohren1983} and a model of monosize carbon
spheres, with optical constants from \citet{Zubko1996}, in order to
interpret the measured slope of Fig.\ref{CorrelWR76} as a dominant size
in the dust size distribution. To do so, we calculated the weighted
extinction coefficients $Q_{ext}(\lambda,a)$ (function of the wavelength
$\lambda$ and the dust sphere radius $a$) as in \citet{Marchenko2003},
the weights being given by $W(\lambda)=T(\lambda)F(\lambda)$, where
$T(\lambda)$ is the transparency of the corresponding filter (V or I)
and $F(\lambda)$ is the energy distribution of WR76 in the optical and
the near infrared \citep{Williams1987}. We then used the fact that the
extinction, $A(\lambda)$, is directly proportional to
$Q_{ext}$. Finally, $A(\lambda)$ is linked to the variations of
magnitude that we interpret as due to an increase of the extinction by
the dust ($A(\lambda) \propto \Delta Mag$ ). The result of our model is
shown Fig.\ref{dustsizeplot} for $a$ ranging from 0.0001 to 0.2 $\mu
m$. Our estimated dust size, corresponding to a slope of $4.47 \pm
0.60$, is $a=0.126 \pm 0.003\, \mu m$. This value of $a$ is quite close
to the estimation of dust size in WR140 (WC7pd+O5) during its periastron
transit in 2001: $a=0.069^{+0.002}_{-0.001}\, \mu m$
\citep{Marchenko2003} but is somewhat lower than other estimations in
WR112 (WC9d+OB?) and WR118 (WC9d): $a=0.4-1.0\, \mu m$
\citep{Chiar2001,Yudin2001,Marchenko2002} .

\section{Conclusions}
The most surprising result of this study is the large fraction of WC8/9
stars that are not variable above the $\sigma \sim$ 0.01 mag level. Even
among the 12 that do vary, it is mostly at a relatively modest level,
$\sigma \lesssim$ 0.02 mag, in contrast to their highly variable WN
cousins, the WN8 stars. We see a significant variation in both V and I
filters for only one object, WR76, which we ascribe to variable
circumstellar dust extinction. We estimated the dominant dust grain size
to be close to 0.1 $\mu$m, which is $\sim$ an order of magnitude smaller
than estimated in two other WC9d stars.

\section*{Acknowledgements}

AFJM is grateful to NSERC (Canada) and FQRNT (Quebec) for financial
assistance.  AZB acknowledges research and travel support from the
Carnegie Institution of Washington through a Vera Rubin Fellowship. We
thank the referee P.M. Williams for useful comments and S.V. Marchenko
for providing a helpful code.

\bibliographystyle{mn2e.bst}

\label{lastpage}
\clearpage

\end{document}